\documentclass{aa}
\usepackage{graphicx}
\def\dg{^{\circ}}
\def\He1211{\mbox{HE\,1211-1707}}
\def\Grw{\mbox{Grw~$+70\dg 8247$}}
\newcommand{\lppr}{\stackrel{<}{\scriptstyle \sim}}
\newcommand{\lappr}{\raisebox{-0.4ex}{$\lppr$}}
\begin{document}
   \title{Stationary components of \ion{He}{I} in strong magnetic
fields -- \\
a tool to identify magnetic DB white dwarfs}

   \author{S. Jordan
          \inst{1}
          \and
          P. Schmelcher\inst{2}
\and W. Becken \inst{2}
          }

   \offprints{S. Jordan}

\institute{Institut f\"ur Theoretische
 Physik und Astrophysik, D-24098 Kiel, Germany\\
\email{jordan@astrophysik.uni-kiel.de}
         \and
Institut f\"ur Physikalische Chemie, Im Neuenheimer Feld 229, D-69120
 Heidelberg, Germany \\
\email{peter@tc.pci.uni-heidelberg.de}
             }

   \date{}

   \abstract{
In only three of the 61 known magnetic white dwarfs helium has been identified unambiguously 
 while 
about  20\%\ of all non-magnetic stars of this class are known to contain
 \ion{He}{I} 
or \ion{He}{II}. 
One reason for this discrepancy is that the identification of peculiar objects as magnetic white
dwarfs is based either on the presence of hydrogen line components in strong magnetic fields --- for 
which atomic data exist since 1984 ---  or the polarization of the corresponding radiation
which has not been measured for many objects.
Until recently, data for \ion{He}{I} data were available  only for magnetic fields below 20\,MG. This changed
with the publication of extensive data by the group in Heidelberg. The corresponding
calculations have now been completed for the energetically lowest five states of singlet and triplet symmetry
for the subspaces with  $|m| \le 3$; selected calculations have been performed for even higher
excitations. In strongly magnetized white dwarfs
only line components are visible whose wavelengths vary  slowly with respect
to the magnetic field, particularly stationary components which have a
wavelength minimum or maximum in the range of the magnetic fields strengths
on the stellar surface. In view of the many ongoing surveys finding white
dwarfs we  want to provide the astronomical community with a tool to identify helium in white dwarfs
for fields up to 5.3\,GG. To this end we present all calculated helium line components whose wavelengths
in the UV, optical, and near IR vary slowly enough with respect to the field strength to produce visible
absorption features.  We also  list all stationary line components in this spectral range. Finally, we find
series of minima and maxima which occur as a result of series of extremal transitions to increasingly higher
excitations. We estimated the limits for 8 series which can possibly give rise to additional 
absorption in white dwarf spectra; one strong absorption feature in GD229 which is yet unexplained
by stationary components is very close to two estimated series limits.
   }

\authorrunning{Jordan, Schmelcher, Becken}
\titlerunning{Stationary components of \ion{He}{I}}
   \maketitle
%                                     Two column figure (place early!)
%______________________________________________ 
   \begin{figure*}
   \centering
   \includegraphics[width=0.85\textwidth]{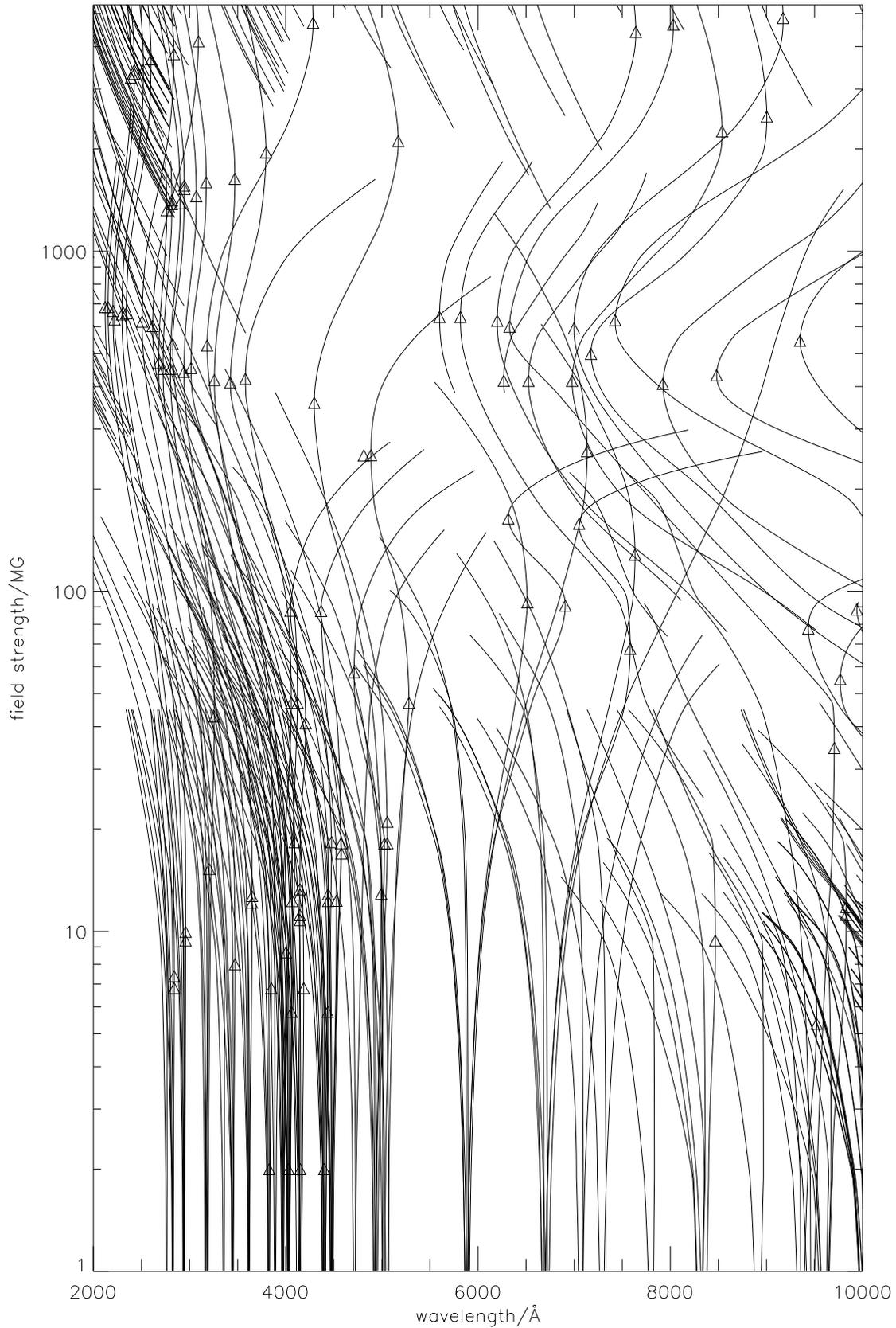}
   \caption{Wavelengths of all calculated stationary line components of \ion{He}{I} 
(minimum or maximum wavelengths are
marked by triangles) and, additionally, of all other
components which vary by 
less than 500\,\AA\ while the magnetic field changes by more than a factor of
two. Note the extremely large number of stationary components
between 300 and 700\,MG corresponding to the field strength of GD\,229
}
              \label{statio}%
    \end{figure*}

%                                     One column figure (place early!)
%______________________________________________ 
   \begin{figure}
   \centering
   \includegraphics[width=0.5\textwidth]{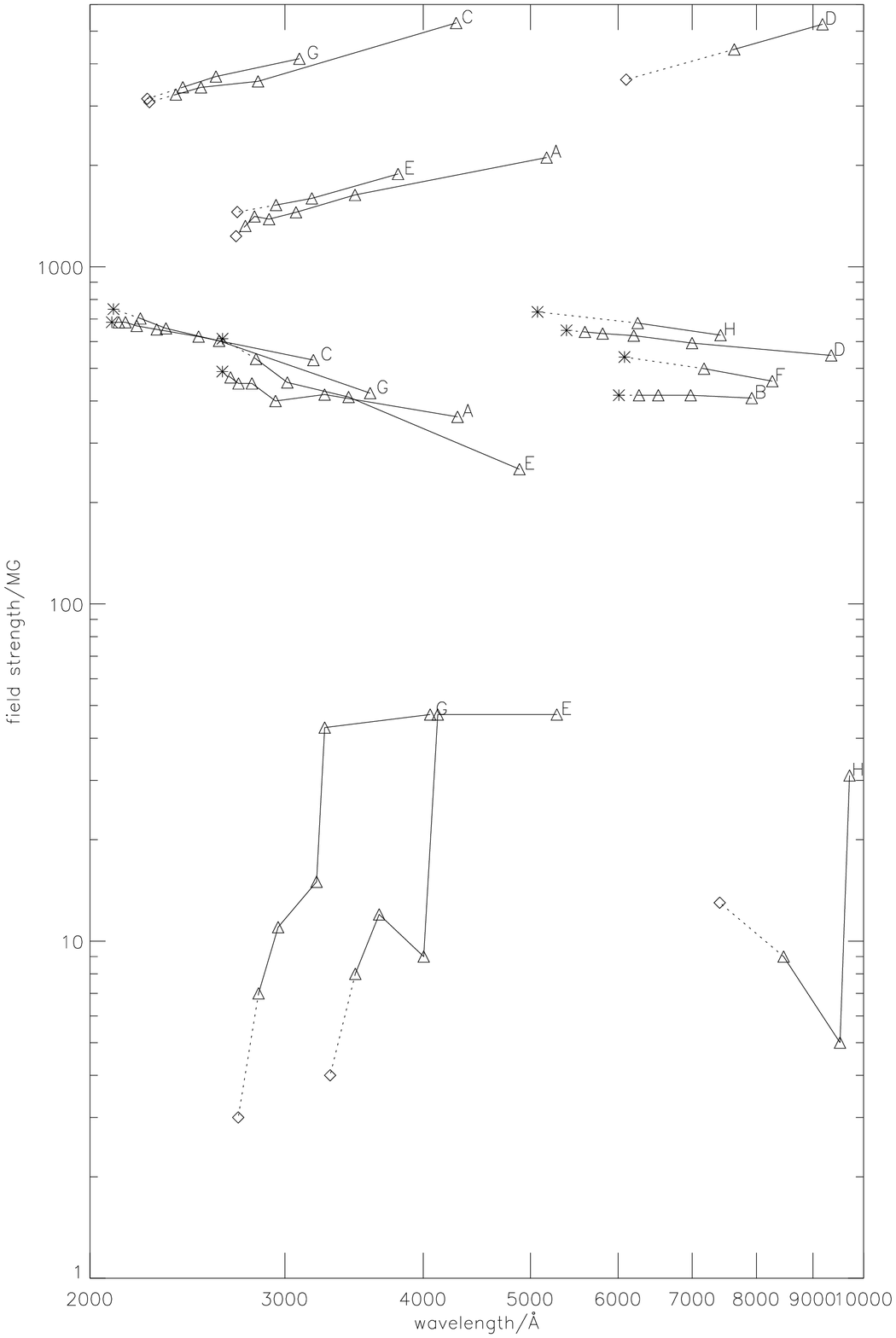}
   \caption{Positions of the stationary components of
the series A$-$H (triangles) and the estimated series 
limits (asterisk for limits of minima, squares for
limits of maxima)
}
              \label{limit}%
    \end{figure}

\section{Introduction}
In about 3-4\%\ of all white dwarfs magnetic fields have been detected
ranging from 2\,kG in 40 Eri B (Fabrika\& Valyavin \cite{FV99}) up to
1\,GG (Schmidt et al.\,\cite{S86}, Latter et al.\,\cite{L87}) in the case of
PG~1031+234. 15 of these 61 objects (Wickramasinghe \&\ Ferrario 
\cite{WF2000}, Jordan \cite{J2001}) have fields in
excess of 100\,MG making them comparable to the typical field
strengths found in millisecond pulsars. Since their magnetic fields
considerably exceed what can be produced in terrestrial laboratories 
they represent cosmic laboratories that probe the behaviour and
properties of atoms and molecules under such extreme conditions
(for reviews of the subject see Ruder et al \cite{Rud94} and Schmelcher \&\ Schweizer \cite{Schm98}).

Until 1984 atomic data for hydrogen were limited to field strengths below
20\,MG (Kemic \cite{K74}) and above $10^{10}$\,G (Garstang \cite{G77}) and it
was a pure  speculation that the unidentified shallow ``Minkowski bands'' 
(Minkowski \cite{M38})  in the spectrum of \Grw\ 
%{\bf{
were due to magnetic fields.  The detection of  circular polarization 
(Kemp \cite{K70}) showed that a magnetic
field exists on this star. The strength, however, could not be determined 
until
%}}
Forster et al. (\cite{F84}) and Henry \&\ O'Connell
(\cite{HO84})  published energy level shifts and transition
probabilities for line components of hydrogen in the intermediate-field
range.

%{\bf{
Particularly%
%}}
,  the so called stationary line components, which go through maxima or
minima as functions of the magnetic field strength lead to significant
absorption structures. Lines whose behavior is
%{\bf{
monotonic and whose wavelengths change significantly with the field
strength
%}}
 are smeared out due to the inhomogeneous magnetic
field over the surface of the white dwarfs.
The identification of the stationary  components of hydrogen 
led to an estimation of
the approximate range of field  strengths covering the stellar surface of
\Grw: 150 to 500\,MG (Greenstein \cite{G84}, Greenstein et al.
\cite{G85}, Angel et al. \cite{A85}, Wunner et al. \cite{W85}) and
identified the star as spectral type DAP.
Later, \Grw\ and other magnetic white dwarfs have been analyzed with
more sophisticated methods using  detailed simulations for the radiative
transfer through magnetized stellar atmospheres (Wickramasinghe \&\ Ferrario \cite{WF88};
%Jordan \cite{J88},
\cite{J89}). However, the basic explanation and 
the approximate field strengths could be inferred from a simple
comparison of the  position of absorption features with the lists
of stationary line components for hydrogen.

Until recently, no reliable atomic data for the intermediate
range of magnetic field strengths existed  for neutral helium. Only in the
case of Feige 7 Achilleos et al. (\cite{A92}) could identify \ion{He}{I}
besides hydrogen in an off-centered dipole field of strength 35\,MG
with an extrapolation of  the Kemic (\cite{K74}) data. On the other hand
several objects -- suspected to be magnetic due to the strange spectral
features and their polarization --- remained unexplained.

Even for the simplest atom i.e. hydrogen an investigation of its
electronic structure is very intricate in the presence of a
strong magnetic field. The degrees of freedom perpendicular and parallel
to the external field cannot be separated but mix strongly. The latter is due
to the competition of the Coulomb and magnetic interaction which possess
different symmetries. Energetically low-lying levels can easily
be obtained only for fields $0.01 > \gamma = B / 2.35 \cdot 10^{9}$ G
(Zeeman and Paschen-Back regime) where perturbation theory applies or for $\gamma \gg 1 $ 
where the dominant magnetic interaction perpendicular to the field
leads to an approximate adiabatic separation of the variables.
In the case of helium the situation is significantly more complicated due to the
electron-electron interaction. First calculations of helium atoms in the intermediate
field regime (Braun et al. \cite{B98}, Jones et al. \cite{Jon97},
Ruder et al. \cite{Rud94} and references therein) either lacked the accuracy
or could access only a few excited states of certain selected symmetries and
for a few selected field strengths. As a consequence a direct comparison with observed spectra from magnetic
white dwarfs was not possible. This changed not before
Becken et al. (\cite{B99}) and  Becken \&\ Schmelcher (\cite{B00})
published a large grid of atomic data for helium at arbitrary magnetic
field strengths. 

The most famous example of a suspected helium rich magnetic white dwarf was GD\,229 
having a large number of strong absorption features in the optical and UV.  With the
help of the newly available data  Jordan et al. (\cite{J98}) could attribute most
of the absorption structures in the spectrum with stationary line components of
neutral helium in a  range of magnetic fields  between 300 and 700\,MG.

The number of magnetic  white dwarfs with clearly identified helium 
lines is still very small: Besides Feige 7 and GD\,229 \ion{He}{I}
could only be identified (Jordan \cite{J2001}) in HE0241-0155
($B\lappr 25$\,MG) and with less certainty in \He1211\ ($20<B< 150$\,MG).

Without reliable atomic data  Reimers et al. (\cite{R98}) have tentatively
identified \ion{He}{I} at a field strength between 20 and 30\,MG
in  HE~0107-0158, HE~0026--2150, and HE~0003--5701.
However, it now turned out that this conclusion could be discarded after applying
our more accurate  atomic data; probably these systems are binary stars of which one component is a 
subdwarf.

Since the publications by Becken et al. (\cite{B99}) and Becken \&\ 
Schmelcher (\cite{B00}) the data for helium in strong magnetic
fields have now been completed for all transitions between the five lowest
energy levels with magnetic quantum numbers $m=-3,-2, -1, 0, +1, +2, +3$ for
both spin singlet and triplet states and for both z-parities with the 
exclusion of the negative z-parity states for $m=3$. To obtain certain desired
stationarities even higher i.e. the sixth and seventh excitations of certain symmetries
have been investigated (see below). Apart from this we have slightly improved the accuracy of our previous
results by enlarging the corresponding basis sets employed in the electronic structure calculations.

The whole data set for \ion{He}{I} comprises of about 2000 line components.
With this paper we want to provide the astronomical
community with an up-to-date overview of all calculated stationary and slowly varying line components of helium
in the regimes of field strengths relevant for magnetic white dwarfs. 
This is particularly useful for an identification of helium-rich magnetic
white dwarfs expected to be found by the large number of ongoing surveys (e.g.
Sloan Digital Sky Survey, York et al. \cite{Y2000}; Hamburg ESO Survey, Wisotzki et al.\ \cite{HES1};
Hamburg Quasar Survey, Hagen et al.\ \cite{HQS1}; Edinburgh Cape Survey, Stobie et al.\ \cite{StobieEC};
Montreal Cambridge Tololo Survey, Lamontagne et al.\cite{LamontagneMCT};
Second Byurakan Survey, Stepanian \cite{StepanianSBS}).

\section{Helium atom in strong magnetic fields}
The  electronic Hamiltonian of the helium atom for fixed nucleus and subjected to a strong magnetic field is given by
\begin{eqnarray}
\cal{H} = & \sum\limits_{i=1}^{2} \left(\frac{1}{2m} {\bf{p}}_i^2 -\frac{e}{2m}
 B L_{z_i} \right.
+ \frac{e^2}{8m} B^2 (x_i^2 + y_i^2) \nonumber \\
&\left. - \frac{2e^2}{|{\bf{r}}_i|}
 - \frac{e}{m} B S_{z_i} \right) + \frac{e^2}{|{\bf{r}}_1-{\bf{r}}_2|}.
\end{eqnarray}

The magnetic field is oriented along the $z-$axis. The respective terms
describe the field-free kinetic energy, the orbital Zeeman term, the diamagnetic interaction,
the Coulomb attraction, the spin Zeeman energies, and  the repulsive
electron-electron interaction. Results beyond the presently applied fixed nucleus approximation
can be obtained via the corresponding scaling relations (Becken \&\ Schmelcher 
\cite{B00}).

The main advantage of our  calculations is the use of an extremely flexible
basis set of anisotropic orbitals with nonlinear variational parameters
$\{\alpha_i,\beta_i\}$ which are adjusted i.e. optimized for 
each magnetic field strength:
\begin{equation}
\Phi_i (\rho,\phi,z) = \rho^{n_{\rho_i}} z^{n_{z_i}} 
 \mbox{exp}\left(-\alpha_i\rho^2
-\beta_iz^2\right) \mbox{exp}\left(i m_i \phi\right)
\end{equation}
where $n_{\rho_i}=|m_i|+2k_i, n_{z_i}= \pi_{z_i} + 2l_i$ with $k_i,l_i = 
0,1,2,\dots$
$\{\alpha_i,\beta_i\}$ are nonlinear variational parameters; $\pi_{z_i}$
is the $z$ parity of the one particle function. 
For the investigation of the electronic structure of the helium atom
we built from the above optimized atomic orbitals the corresponding
symmetry adapted two-particle configurations and represent the Hamiltonian
matrix in these configurations (the typical dimension of the full matrices
is $5000$). To this end a fast and accurate evaluation
of the corresponding electron-electron matrix elements is crucial. A number
of advanced analytical as well as numerical techniques have been employed
to achieve this goal. For details on the latter we refer the reader to
Becken et al. (\cite{B99}) and Becken \& Schmelcher (\cite{B00}).
To obtain the eigenenergies and eigenfunctions of the atom a full configuration
interaction approach is used.  Due to the nonorthogonality
of the above basis set the latter leads to a generalized eigenvalue problem which can be
diagonalized using standard library routines. We remark that it is only due to the
above-indicated extensive developmental work that a series
production of accurate atomic data for the helium atom in the presence of the
strong field became possible.

%{\bf{
In the presence of a magnetic field the total angular momentum of the atom is
not conserved and provides therefore not a good quantum number.
The only remaining spatial constant of motion is the projection of the total angular momentum
$\sum_i L_{z_i}$ onto the magnetic field axis (quantum number $M$). Furthermore 
the total $z-$parity ($\Pi_z: z_i \rightarrow -z_i$) represents a symmetry of the
atom (total parity is also conserved but can be obtained from the previous symmetries).
The spin symmetries ($S=1,3$ for singlet and triplet, respectively)
are the same as in the absence of a magnetic field.
We therefore use the spectroscopic notation $n ^{S}M^{\Pi_z}$ where $n$ indicates 
the energetical degree of excitation.
%}}

\section{Stationary components}
Stationary components remain visible i.e. may lead to absorption edges in the
observed spectra even when the magnetic field strength varies 
considerably over the surface of a white dwarf.

Table\,\ref{statiotaba} lists all 101  singlet and triplet stationary line components
of \ion{He}{I} with $|m| \le 3$ and for both z-parities (for $|m|=3$ only the positive 
$z$ parity has been considered). Included are those transitions whose wavelengths belong
to the regime from the FUV ($>900$\,\AA)  to the near
infrared ($<10000$\,\AA) and run through minima or maxima with respect to the varying
magnetic field strength.  No stationary components were found in the regime 900-2100\,\AA.
The white dwarf with the highest field strength discovered so far, PG\,1031+234,
has a maximum field strength of about 1\,GG; to be on the safe side concerning possible higher 
field strengths we have listed in Table\,\ref{statiotaba} all stationary components up to 5.3\,GG.

Stationarity is not a necessary condition for line components to produce visible
absorption features; therefore we also looked for  components whose
wavelengths are not stationary but do not change by more than 500\,\AA\ for a variation of the
magnetic field by more than a factor of two (corresponding to the spread of
centered dipole fields from the magnetic pole to the equator). We expect the latter
to be also candidates for observable absorption edges.
In Fig.\,\ref{statio} the wavelengths of these components are shown together
with the stationary ones for varying magnetic field strength.
The 
%{\bf{
uncertainties of the   calculated wavelengths are smaller
%}}
than a few \AA. However, since the lines are
interpolated on a grid for 20 different magnetic field strengths the 
typical accuracy with respect to the position of the minima or maxima is about 10\,\AA;
%{\bf{
only in a few cases the error can go up to 50\,\AA. These values were estimated
by comparing the results of different interpolation schemes.
%}}
In the case of GD~229 26 stationary components could be identified in the optical and UV
spectrum in Jordan et al. \cite{J98}. Such a rich spectrum is probably an
exception since the number of stationary line components
in the range of magnetic fields  between 300 and 700\,MG is much larger
than in any other comparable range of magnetic field strengths.
On the other hand we expect a priori that in the intermediate regime of field strengths a
severe rearrangement of the electronic wave functions takes place and therefore
a strong dependence of the energy levels on the field strength has also to be expected (the above
mentioned interval 300 to 700\,MG is contained in the intermediate regime).
Since the publication by Jordan et al. (\cite{J98}), which analyzed line components for $|m| \le 1$,
the investigations on the electronic eigenstates
of the helium atom in a magnetic field proceeded significantly and in particular
a large number of higher excited states covering new symmetries have been studied (see above).
As a consequence seven new extrema have been obtained in the interval 
300 to 700\,MG. They possess the wavelengths $2204, 2679, 2724, 5600, 5810, 6267$ \AA\ and
$6524$ \AA. With the exception of the wavelength $5810$ \AA\ these new extrema also match nicely
with the absorption features of the magnetic white dwarf GD~229 which confirms our previous
conclusion of strong evidence for helium on this white dwarf. 

However, not everything is completely clarified concerning the spectrum of GD~229
even within the framework of its interpretation in terms of stationary transitions. 
The strong absorption feature at $4000-4200$\,\AA\ is met 'only' by the 
stationary transition $2^10^+ \rightarrow 2^10^-$  which cannot
account for absorption at $\lambda<4296$\AA. Furthermore the absorption edge
at approximately $5280$\,\AA\ has no stationary counterpart even within the
significantly enlarged data set of atomic calculations. A careful look
at the stationary atomic line components particularly relevant for the interpretation of
GD~229 (300 to 700\,MG), however, reveals that these stationarities occur in terms of series of transitions
to increasingly higher excitations. The series (see Table\,\ref{statiotaba}) are

\begin{eqnarray}
\label{seriesid}
A:~~2^10^+ &\rightarrow& n^10^-, n\ge2\\ \nonumber
B:~~3^10^+ &\rightarrow& n^10^-, n\ge3\\ \nonumber
C:~~1^30^+ &\rightarrow& n^30^-, n\ge2\\ \nonumber
D:~~2^30^+ &\rightarrow& n^30^-, n\ge3\\ \nonumber
E:~~2^10^+ &\rightarrow& n^1(-1)^+, n\ge2\\ \nonumber
F:~~3^10^+ &\rightarrow& n^1(-1)^+, n\ge4\\ \nonumber
G:~~1^30^+ &\rightarrow& n^3(-1)^+, n\ge2\\ \nonumber
H:~~2^30^+ &\rightarrow& n^3(-1)^+, n\ge4\\ \nonumber
\end{eqnarray}

Although an ab initio electronic structure investigation can always reliably calculate only a finite number
of excitations it suggests itself that the complete above series for arbitrary $n$ lead to stationarities.
Within our atomic calculations we could show the stationary character of the corresponding transitions
for $n \le 7$ for the states of the $^10^-,^30^-$ subspaces and for $n\le 5$ for the subspaces
$^1(-1)^+,^3(-1)^+$. Let us explain the above statement in more
detail for the series D i.e. $2^30^+ \rightarrow n^30^-$. The corresponding stationary wavelengths
are $9348, 6998, 6199, 5810, 5600$\AA\ for $n=3-7$. This shows exemplary that the series of
stationarities converges towards a series limit at which there is an infinite
accumulation of stationarities. This property seems to hold for all of the above
series and is an amazing fact of the electronic structure of the atom in the considered particular part of
the intermediate regime of field strengths.
The presence of the infinite series of stationaries can be further
elucidated by observing that the energies of the initial states
$2^10^+,3^10^+,1^30^+,2^30^+,2^10^+,3^10^+,1^30^+,2^30^+$ of the series A$-$H
show already the corresponding minima and maxima responsible for the series
of stationarities. Due to the dominating dependence of the energies of these
low-lying states on the field strength they leave their fingerprint on any
transition to higher excited states which as a consequence become stationary
for a certain field strength.
>From the available atomic data we can now estimate the
wavelengths of the series limits by using the empirical law that the wavelengths
become smaller by a factor of two for successive transitions for high $n$. 
 For the series $A-H$ the limiting wavelengths are listed in
Table\,\ref{statiolimits}.  Fig.\,\ref{limit} shows that this
rule applies well for all stationary components above 100\,MG.
 We remark that the series limits of $F$ and $H$ are particularly crude estimates. Nevertheless these estimates
nicely fit with the position of major absorption edges in the spectrum of GD~229 and in particular
the estimated series limits of $D$ and $H$ (5390 and 5074\,\AA, respectively) 
come very close to the so far unexplained absorption edge at approximately 
$5300$\AA.  In view of the above one can conjecture that the accumulation of stationarities might be
responsible for certain observed features of the spectrum. Future calculations on the oscillator strengths
of the stationary line components will certainly help to further clarify this problem. 
The role of bound-free transitions for the helium atom in the presence of the strong field and
in particular its impact on observable spectra from white dwarf atmospheres is equally an open question. The
corresponding investigations require however major theoretical and computational developments.
Generally it is assumed that the higher excited states are strongly de-populated by the interaction
of close atoms in the high density atmospheres of the white dwarfs.

Of course we cannot strictly exclude that further relevant
stationary line components arise due to the transitions involving other symmetry subspaces
and states than those discussed above. However according to the previous discussion the
regularity with respect to the emergence of stationary transitions particularly in the field regime
300 to 700\,MG suggests that all major components have been addressed. Additionally one has to keep in mind
that the level spacing decreases rapidly with increasing magnetic quantum numbers for the states
involved and therefore the wavelength of the corresponding transitions become increasingly larger
and are not of relevance to the observed spectral range. For example, the series of circular 
polarized transitions $1^1(-2)^+ \rightarrow n^1(-3)^+, n = 1-5$ possesses stationarities for the wavelengths
$21400, 13000, 11000, 10900$ and $10000$\AA, respectively, which excludes it from the
spectral regime considered here. Even higher excitations $n \ge 6$ will not be significantly 
below $10000$\AA. These arguments hold for both singlet and triplet excitations: with increasing
degree of excitation the singlet and triplet splitting decreases rapidly. A similar argument
holds also for the linear polarized stationary transitions like e.g. $3^1(-2)^+ \rightarrow n^1(-2)^-, n\ge 2$.
This tendency is even enforced for the corresponding transitions involving the $|m| \ge 4$ subspaces. 
We emphasize that the above arguments are valid in the regime of field strengths considered here
i.e. for $B \le 5.3$\,GG. In the high-field regime $B \ge 10$\,GG severe changes can be observed.

Some remarks are in order concerning a recent work which suggests an alternative interpretation
of the spectrum of GD~229. Jones et al (\cite{Jon99}) used their results on helium calculations
in strong fields in order to conclude that both \ion{He}{I} as well as \ion{He}{II} could 
contribute to the observed spectrum. However the authors dealt with a significantly smaller set of
atomic data and therefore missed many of the stationarities given in the present work. 
Moreover their explanation is based on the existence of \ion{He}{II} in the atmosphere of
GD~229 which is a critical issue since the effective temperature (about 16000\,K according to
Schmidt et al. \cite{Schmidt90})
is generally not assumed to be large
enough in order to expect \ion{He}{II}; in non-magnetic white dwarfs models predict \ion{He}{II}
to occur at $T_{rm eff}$ above about 28000\,K. To allow, nevertheless, for the existence of
\ion{He}{II} a double excitation process via a single UV photon followed by a subsequent
radiationless autoionizing process is suggested. To the authors of the present work
the efficiency of this mechanism is very questionable.

\section{Outlook}
With the data presented in this work it will be possible for  observational astronomers to
perform a first identification and get evidence for helium in 
magnetic white dwarfs. This is particularly important due to the large number of magnetic 
objects being discovered by many (ongoing) surveys to find
stellar and extragalactic objects. A positive identification must be
based on the simultaneous assignment of slowly varying or stationary
line components in a realistic range of magnetic field strengths.
Although this on its own does not allow a prediction of the strengths of the
absorption features it provides a good starting point for the parameters relevant to
subsequent numerical calculations of the polarized radiation of magnetic white dwarfs.
The latter can be directly compared to the spectra and polarization measurements.
We are currently working on the inclusion of the energies, wavelengths, and transition
probabilities for all calculated line components for neutral helium
in arbitrary magnetic fields into our models for the radiative transfer. 

>From our comprehensive data set it became clear that GD\,229 with
its large number of absorption features is a very special object since the regime of field strengths covering
its surface ($\approx$ 300-700\,MG) contains a stronger accumulation of stationary transitions of
\ion{He}{I} than any other equally large interval of field strengths.

\begin{acknowledgements}
The Deutsche Studienstiftung and the Deutsche Forschungsgemeinschaft (Schm 885/7-1 \& Ko 738/7-1)
are gratefully acknowledged for financial support.
\end{acknowledgements}

%
%________________________________________________________________

%__________________________________________________ One column table
   \begin{table}
      \caption[]{ Line components of \ion{He}{I} which are
stationary at wavelengths between $900$ and $10000$\,\AA\ for magnetic fields
below 5.3\,GG  sorted by the wavelength
of the minimum or maximum, respectively. Note that the wavelengths
(and corresponding field strengths) of the minima and maxima are interpolated
in a relatively crude grid for 20 different field strengths so that
the  accuracy of the wavelengths varies between about 10 and 50\,\AA.
The membership to the series A$-$H (see Eq.\,\ref{seriesid}) is also
indicated.}
         \label{statiotaba}
\begin{flushleft}
\scriptsize
    \begin{tabular}{lrrrccl}\hline
\noalign{\smallskip}
\multicolumn{2}{r}{$B$/} & \multicolumn{1}{c}{$\lambda/$\AA} & zero field trans.&transition & type \\
\multicolumn{2}{r}{MG} & & &&\\
\noalign{\smallskip}
\hline
\noalign{\smallskip}
%   685.1 &  2122.5 &$2^3 S_0\hspace{-0.6ex}\rightarrow\hspace{-0.6ex}6^3 P_0$&  $1^3 0^+ \hspace{-0.6ex}\rightarrow\hspace{-0.6ex} 7^3 0^-$& min\\
C: &  685 &  2123 &$2^3 S_0    \rightarrow 6^3 P_0    $&  $1^3 0^+ \rightarrow  7^3 0^- $ & min\\
C: &  685 &  2153 &$2^3 S_0    \rightarrow 5^3 F_0    $&  $1^3 0^+ \rightarrow 6^3 0^- $  & min\\
C: &  668 &  2204 &$2^3 S_0    \rightarrow 5^3 P_0    $&  $1^3 0^+ \rightarrow 5^3 0^- $  & min\\
G: &  703 &  2221 &$2^3 S_0    \rightarrow 5^3 P_{-1} $&  $1^3 0^+ \rightarrow 5^3 (-1)^+$& min\\
C: &  652 &  2298 &$2^3 S_0    \rightarrow 4^3 F_0    $&  $1^3 0^+ \rightarrow 4^3 0^-$   & min\\
G: &  657 &  2342 &$2^3 S_0    \rightarrow 4^3 F_{-1} $&  $1^3 0^+ \rightarrow 4^3 (-1)^+$& min\\
C: & 3243 &  2391 &$2^3 S_0    \rightarrow 5^3 P_0    $&  $1^3 0^+ \rightarrow 5^3 0^-$   & max\\
G: & 3409 &  2426 &$2^3 S_0    \rightarrow 5^3 P_{-1} $&  $1^3 0^+ \rightarrow 5^3 (-1)^+$& max\\
C: &  621 &  2507 &$2^3 S_0    \rightarrow 4^3 P_0    $&  $1^3 0^+ \rightarrow 3^3 0^-$   & min\\
C: & 3406 &  2519 &$2^3 S_0    \rightarrow 4^3 F_0    $&  $1^3 0^+ \rightarrow 4^3 0^-$   & max\\
G: & 3668 &  2599 &$2^3 S_0    \rightarrow 4^3 F_{-1} $&  $1^3 0^+ \rightarrow 4^3 (-1)^+$& max\\
G: &  603 &  2616 &$2^3 S_0    \rightarrow 4^3 P_{-1} $&  $1^3 0^+ \rightarrow 3^3 (-1)^+$& min\\
A: &  470 &  2679 &$2^1 S_0    \rightarrow 6^1 F_0    $&  $2^1 0^+ \rightarrow 7^1 0^-$   & min\\
A: &  451 &  2724 &$2^1 S_0    \rightarrow 5^1 P_0    $&  $2^1 0^+ \rightarrow 6^1 0^-$   & min\\
A: & 1322 &  2763 &$2^1 S_0    \rightarrow 6^1 F_0    $&  $2^1 0^+ \rightarrow 7^1 0^-$   & max\\
A: &  451 &  2801 &$2^1 S_0    \rightarrow 5^1 F_0    $&  $2^1 0^+ \rightarrow 5^1 0^-$   & min\\
A: & 1411 &  2816 &$2^1 S_0    \rightarrow 5^1 P_0    $&  $2^1 0^+ \rightarrow 6^1 0^-$   & max\\
E: &  533 &  2825 &$2^1 S_0    \rightarrow 5^1 F_{-1} $&  $2^1 0^+ \rightarrow 5^1(-1)^+$ & min\\
C: & 3549 &  2837 &$2^3 S_0    \rightarrow 4^3 P_0    $&  $1^3 0^+ \rightarrow 3^3 0^-$   & max\\
G: &    7 &  2840 &$2^3 S_0    \rightarrow 6^3 P_{-1} $&  $1^3 0^+ \rightarrow 7^3 (-1)^+$& max\\
A: & 1385 &  2903 &$2^1 S_0    \rightarrow 5^1 F_0    $&  $2^1 0^+ \rightarrow 5^1 0^-$   & max\\
A: &  400 &  2942 &$2^1 S_0    \rightarrow 4^1 P_0    $&  $2^1 0^+ \rightarrow 4^1 0^-$   & min\\ 
E: & 1525 &  2945 &$2^1 S_0    \rightarrow 5^1 F_{-1} $&  $2^1 0^+ \rightarrow 5^1(-1)^+$ & max\\
G: &   11 &  2958 &$2^3 S_0    \rightarrow 5^3 P_{-1} $&  $1^3 0^+ \rightarrow 5^3 (-1)^+$& max\\
E: &  454 &  3015 &$2^1 S_0    \rightarrow 4^1 P_{-1} $&  $2^1 0^+ \rightarrow 4^1(-1)^+$ & min\\
A: & 1451 &  3070 &$2^1 S_0    \rightarrow 4^1 P_0    $&  $2^1 0^+ \rightarrow 4^1 0^-$   & max\\
G: & 4142 &  3092 &$2^3 S_0    \rightarrow 4^3 P_{-1} $&  $1^3 0^+ \rightarrow 3^3 (-1)^+$& max\\
E: & 1596 &  3173 &$2^1 S_0    \rightarrow 4^1 P_{-1} $&  $2^1 0^+ \rightarrow 4^1(-1)^+$ & max\\
C: &  529 &  3183 &$2^3 S_0    \rightarrow 3^3 P_0    $&  $1^3 0^+ \rightarrow 2^3 0^-$   & min\\
G: &   15 &  3204 &$2^3 S_0    \rightarrow 4^3 F_{-1} $&  $1^3 0^+ \rightarrow 4^3 (-1)^+$& max\\
A: &  418 &  3258 &$2^1 S_0    \rightarrow 4^1 F_0    $&  $2^1 0^+ \rightarrow 3^1 0^-$   & min\\
G: &   43 &  3259 &$2^3 S_0    \rightarrow 4^3 P_{-1} $&  $1^3 0^+ \rightarrow 3^3 (-1)^+$& max\\
E: &  411 &  3425 &$2^1 S_0    \rightarrow 4^1 F_{-1} $&  $2^1 0^+ \rightarrow 3^1(-1)^+$ & min\\
A: & 1634 &  3471 &$2^1 S_0    \rightarrow 4^1 F_0    $&  $2^1 0^+ \rightarrow 3^1 0^-$   & max\\
E: &    8 &  3473 &$2^1 S_0    \rightarrow 6^1 F_{-1} $&  $2^1 0^+ \rightarrow 7^1(-1)^+$ & max\\
G: &  422 &  3582 &$2^3 S_0    \rightarrow 3^3 P_{-1} $&  $1^3 0^+ \rightarrow 2^3 (-1)^+$& min\\
E: &   12 &  3650 &$2^1 S_0    \rightarrow 5^1 F_{-1} $&  $2^1 0^+ \rightarrow 5^1(-1)^+$ & max\\
E: & 1885 &  3796 &$2^1 S_0    \rightarrow 4^1 F_{-1} $&  $2^1 0^+ \rightarrow 3^1(-1)^+$ & max\\
   &  2 &  3830 &$2^3 P_{-1} \rightarrow 6^3 D_{-2} $&  $1^3 (-1)^+ \rightarrow 5^3 (-2)^+$& max\\
   &    7 &  3851 &$2^3 P_0    \rightarrow 6^3 D_{-1} $&  $1^3 0^- \rightarrow 5^3 (-1)^- $& max\\
E: &    9 &  4004 &$2^1 S_0    \rightarrow 4^1 P_{-1} $&  $2^1 0^+ \rightarrow 4^1(-1)^+$ & max\\
   &  2 &  4030 &$2^3 P_{-1} \rightarrow 6^3 D_{-2} $&  $1^3 (-1)^+ \rightarrow 4^3 (-2)^+$& max\\
   & 88 &  4050 &$5^3 S_0    \rightarrow 2^3 P_{+1} $&  $6^3 0^+ \rightarrow 1^3 (+1)^+$& min\\
G: &   47 &  4058 &$2^3 S_0    \rightarrow 3^3 P_{-1} $&  $1^3 0^+ \rightarrow 2^3 (-1)^+$& max\\
   &  6 &  4061 &$2^3 P_0    \rightarrow 5^3 G_{-1} $&  $1^3 0^- \rightarrow 4^3 (-1)^-$& max\\
   & 12 &  4060 &$2^3 P_{-1} \rightarrow 5^3 D_{-2} $&  $1^3 (-1)^+ \rightarrow 3^3 (-2)^+$& max\\
   & 18 &  4094 &$2^3 P_0    \rightarrow 5^3 D_{-1} $&  $1^3 0^- \rightarrow 3^3 (-1)^-$& max\\
E: &   47 &  4123 &$2^1 S_0    \rightarrow 4^1 F_{-1} $&  $2^1 0^+ \rightarrow 3^1(-1)^+$ & max\\
   & 13 &  4147 &$5^3 S_0    \rightarrow 2^3 P_{+1} $&  $6^3 0^+ \rightarrow 1^3 (+1)^+$& min\\
   &  2 &  4150 &$2^1 P_{-1} \rightarrow 6^1 D_{-2} $&  $1^1(-1)^+ \rightarrow 5^1(-2)^+$& max\\
  &   7 &  4187 &$2^1 P_0    \rightarrow 6^1 D_{-1} $&  $1^1 0^- \rightarrow 5^1(-1)^-$& max\\
  &  41 &  4208 &$5^3 S_0    \rightarrow 2^3 P_{+1} $&  $6^3 0^+ \rightarrow 1^3 (+1)^+$& max\\
C:&  5288 &  4284 &$2^3 S_0    \rightarrow 3^3 P_0    $&  $1^3 0^+ \rightarrow 2^3 0^-$   &max\\
  & 359 &  4296 &$2^1 S_0    \rightarrow 3^1 P_0    $&  $2^1 0^+ \rightarrow 2^1 0^-$   & min\\
A:  &  88 &  4368 &$4^3 D_0    \rightarrow 2^3 P_{+1} $&  $5^3 0^+ \rightarrow 1^3 (+1)^+$& min\\
  &   2 &  4400 &$2^1 P_{-1} \rightarrow 5^1 G_{-2} $&  $1^1(-1)^+ \rightarrow 4^1(-2)^+$& max\\
  &   5 &  4442 &$2^1 P_0    \rightarrow 5^1 G_{-1} $&  $1^1 0^- \rightarrow 4^1(-1)^-$& max\\
  &  13 &  4441 &$2^1 P_{-1} \rightarrow 5^1 D_{-2} $&  $1^1(-1)^+ \rightarrow 3^1(-2)^+$& max\\
  &  18 &  4477 &$2^1 P_0    \rightarrow 5^1 D_{-1} $&  $1^1 0^- \rightarrow 3^1(-1)^-$& max\\
  &  12 &  4528 &$2^3 P_{-1} \rightarrow 4^3 D_{-2} $&  $1^3 (-1)^+ \rightarrow 2^3 (-2)^+$& max\\
  &  20 &  4574 &$2^3 P_0    \rightarrow 4^3 D_{-1} $&  $1^3 0^- \rightarrow 2^3 (-1)^-$& max\\
\noalign{\smallskip}
\hline
    \end{tabular}

    \end{flushleft}
   \end{table}
\normalsize

\addtocounter{table}{-1}
   \begin{table}
      \caption[]{ continued }
\begin{flushleft}
         \label{statiotab}
\scriptsize
    \begin{tabular}{lrrccl}\hline
\noalign{\smallskip}
\multicolumn{2}{r}{$B$/} & \multicolumn{1}{c}{$\lambda/$\AA} & zero field trans.&transition & type \\
\multicolumn{2}{r}{MG}   &&\\
\noalign{\smallskip}
\hline
\noalign{\smallskip}
  &  17 &  4578 &$4^3 D_0    \rightarrow 2^3 P_{+1} $&  $5^3 0^+ \rightarrow 1^3 (+1)^+$& max\\
  &  58 &  4711 &$4^1 D_0    \rightarrow 2^1 P_{+1} $&  $6^1 0^+ \rightarrow 1^1(+1)^+$& min\\
E:&   251 &  4812 &$2^1 S_0    \rightarrow 3^1 P_{-1} $&  $2^1 0^+ \rightarrow 2^1(-1)^+$ & min\\
  &  13 &  4993 &$2^1 P_{-1} \rightarrow 4^1 D_{-2} $&  $1^1(-1)^+ \rightarrow 2^1(-2)^+$& max\\
  &  18 &  5026 &$4^1 D_0    \rightarrow 2^1 P_{+1} $&  $6^1 0^+ \rightarrow 1^1(+1)^+$& max\\
  &  21 &  5056 &  $2^1 P_0    \rightarrow 4^1 D_{-1} $&  $1^1 0^- \rightarrow 2^1(-1)^-$& max\\
A:&  2109 &  5171 &$2^1 S_0    \rightarrow 3^1 P_0    $&  $2^1 0^+ \rightarrow 2^1 0^-$   & max\\
E:&   47 &  5282 &$2^1 S_0    \rightarrow 3^1 P_{-1} $&  $2^1 0^+ \rightarrow 2^1(-1)^+$ & max\\
D:&   641 &  5600 &$3^3 S_0    \rightarrow 6^3 P_0    $&  $2^3 0^+ \rightarrow 7^3 0^-$   & min\\
D:&   634 &  5810 &$3^3 S_0    \rightarrow 5^3 F_0    $&  $2^3 0^+ \rightarrow 6^3 0^-$   & min\\
D:&   625 &  6199 &$3^3 S_0    \rightarrow 5^3 P_0    $&  $2^3 0^+ \rightarrow 5^3 0^-$   & min\\
B:&   416 &  6267 &$3^1 S_0    \rightarrow 6^1 F_0    $&  $3^1 0^+ \rightarrow 7^1 0^-$   & min\\
  & 164 &  6314 &$3^3 D_0    \rightarrow 2^3 P_{+1} $&  $3^3 0^+ \rightarrow 1^3 (+1)^+$& min\\
H:&   681 &  6250 &$3^3 S_0    \rightarrow 5^3 P_{-1} $&  $2^3 0^+ \rightarrow 5^3 (-1)^+$& min\\
  &  93 &  6513 &$2^3 P_{-1} \rightarrow 3^3 D_{-2} $&  $1^3 (-1)^+ \rightarrow 1^3 (-2)^+$& max\\
B:&   416 &  6524 &$3^1 S_0    \rightarrow 5^1 P_0    $&  $3^1 0^+ \rightarrow 6^1 0^-$   & min\\
  &  91 &  6907 &$3^3 D_0    \rightarrow 2^3 P_{+1} $&  $3^3 0^+ \rightarrow 1^3 (+1)^+$& max\\
B:&   416 &  6978 &$3^1 S_0    \rightarrow 5^1 F_0    $&  $3^1 0^+ \rightarrow 5^1 0^-$   & min\\
D:&   594 &  6998 &$3^3 S_0    \rightarrow 4^3 F_0    $&  $2^3 0^+ \rightarrow 4^3 0^-$   & min\\
  & 158 &  7051 &$4^1 S_0    \rightarrow 6^1 P_{+1} $&  $5^1 0^+ \rightarrow 9^1(+1)^+$& min\\
  & 258 &  7135 &$2^3 P_0    \rightarrow 3^3 D_{-1} $&  $1^3 0^- \rightarrow 1^3 (-1)^-$& max\\
F:&   499 &  7175 &$3^1 S_0    \rightarrow 5^1 F_{-1} $&  $3^1 0^+ \rightarrow 5^1(-1)^+$ & min\\
H:&   627 &  7426 &$3^3 S_0    \rightarrow 4^3 F_{-1} $&  $2^3 0^+ \rightarrow 4^3 (-1)^+$& min\\
  &  68 &  7582 &$4^1 S_0    \rightarrow 6^1 P_{+1} $&  $5^1 0^+ \rightarrow 9^1(+1)^+$& max\\
  & 128 &  7632 &$2^1 P_{-1} \rightarrow 3^1 D_{-2} $&  $1^1(-1)^+ \rightarrow 1^1(-2)^+$& max\\
D:&  4416 &  7641 &$3^3 S_0    \rightarrow 5^3 P_0    $&  $2^3 0^+ \rightarrow 5^3 0^-$   & max\\
B:&   408 &  7923 &$3^1 S_0    \rightarrow 4^1 P_0    $&  $3^1 0^+ \rightarrow 4^1 0^-$   & min\\
H:&  5029 &  8032 &$3^3 S_0    \rightarrow 5^3 P_{-1} $&  $2^3 0^+ \rightarrow 5^3 (-1)^+$& max\\
H:&     9 &  8467 &$3^3 S_0    \rightarrow 6^3 P_{-1} $&  $2^3 0^+ \rightarrow 7^3 (-1)^+$& max\\
F:&   458 &  8270 &$3^1 S_0    \rightarrow 4^1 P_{-1} $&  $3^1 0^+ \rightarrow 4^1(-1)^+$ & min\\
B:&  2253 &  8539 &$3^1 S_0    \rightarrow 5^1 F_0    $&  $3^1 0^+ \rightarrow 5^1 0^-$   & max\\
F:&  2489 &  9003 &$3^1 S_0    \rightarrow 5^1 F_{-1} $&  $3^1 0^+ \rightarrow 5^1(-1)^+$ & max\\
D:&  5240 &  9180 &$3^3 S_0    \rightarrow 4^3 F_0    $&  $2^3 0^+ \rightarrow 4^3 0^-$   & max\\
D:&   546 &  9348 &$3^3 S_0    \rightarrow 4^3 P_0    $&  $2^3 0^+ \rightarrow 3^3 0^-$   & min\\
  &  78 &  9434 &$3^3 D_0    \rightarrow 5^3 F_0    $&  $3^3 0^+ \rightarrow 6^3 0^-$& min\\
H:&     5 &  9521 &$3^3 S_0    \rightarrow 5^3 F_{-1} $&  $2^3 0^+ \rightarrow 6^3 (-1)^+$& max\\
H:&    31 &  9711 &$3^3 S_0    \rightarrow 5^3 P_{-1} $&  $2^3 0^+ \rightarrow 5^3 (-1)^+$& max\\
  &  56 &  9730 &$3^1 D_0    \rightarrow 5^1 P_0    $&  $4^1 0^+ \rightarrow 6^1 0^-$& min\\
F:&    11 &  9829 &$3^1 S_0    \rightarrow 6^1 F_{-1} $&  $3^1 0^+ \rightarrow 7^1(-1)^+$ & max\\
  &  88 &  9939 &$3^3 D_0    \rightarrow 5^3 F_{-1} $&  $3^3 0^+ \rightarrow 6^3 (-1)^+$& min\\
\noalign{\smallskip}
\hline
    \end{tabular}
\end{flushleft}
   \end{table}

   \begin{table}
      \caption[]{Crude estimates for the  series limits calculated
under the assumption that the differences in wavelengths divide by two for
successive higher series members
 }
\begin{flushleft}
         \label{statiolimits}
\scriptsize
    \begin{tabular}{lrrccl}\hline
\noalign{\smallskip}
\multicolumn{2}{r}{$B$/} & \multicolumn{1}{c}{$\lambda/$\AA} &transition & type \\
\multicolumn{2}{r}{MG}  \\
\noalign{\smallskip}
\hline
\noalign{\smallskip}
A: &  490 &  2634 &  $2^1 0^+ \rightarrow \infty^1 0^-$& min\\
A: & 1230 &  2710 &  $2^1 0^+ \rightarrow \infty^1 0^-$& max\\
B: & 420 &  6010  &  $3^1 0^+ \rightarrow \infty^1 0^-$& min\\
C: &  690 &  2093 &  $1^3 0^+ \rightarrow \infty^3 0^- $& min\\
C: & 3080 &  2263 &  $1^3 0^+ \rightarrow \infty^3 0^-$& max\\
D: &  650 &  5390 &  $2^3 0^+ \rightarrow \infty^3 0^-$& min\\
D: & 3590 &  6102 &  $2^3 0^+ \rightarrow \infty^3 0^-$& max\\
E: & 1450 &  2717 &  $2^1 0^+ \rightarrow \infty^1(-1)^+$& max\\
E: &  610 &  2635 &  $2^1 0^+ \rightarrow \infty^1(-1)^+$& min\\
E: &    4 &  3296 &  $2^1 0^+ \rightarrow \infty^1(-1)^+$& max\\
F: &  540 &  6080 &  $3^1 0^+ \rightarrow \infty^1(-1)^+$& min\\
G: & 3150 &  2253 &  $1^3 0^+ \rightarrow \infty^3 (-1)^+$& max\\
G: &  735 &  2100 &  $1^3 0^+ \rightarrow \infty^3 (-1)^+$& min\\
G: &    3 &  2722 &  $1^3 0^+ \rightarrow \infty^3 (-1)^+$& max\\
H: &  740 &  5074 &  $2^3 0^+ \rightarrow \infty^3 (-1)^+$& min\\
H: &   13 &  7413 &  $2^3 0^+ \rightarrow \infty^3 (-1)^+$& max\\
\noalign{\smallskip}
\hline
    \end{tabular}
\end{flushleft}
   \end{table}
\end{document}